\documentclass[twoside]{dis09}
\usepackage[latin1]{inputenc}
\usepackage[dvips]{graphicx,epsfig,color}
\usepackage{wrapfig,rotating}
\usepackage{amssymb,amsmath,array}

\usepackage{psfrag}

\begin{document}

\newcommand{\zbar}{\bar{z}}
\newcommand{\odd}{\mathbb{O}}
\newcommand{\pom}{\mathbb{P}}
\newcommand{\tmin}{t_{\rm min}}
\newcommand{\smin}{s_{\rm min}}
\newcommand{\gev}{{\rm GeV}}

\title{Pomeron Odderon interference in production of\\ $\pi^+\pi^-$ pairs at LHC}

\author{B. Pire$^1$, F. Schwennsen$^{1,2}$, L. Szymanowski$^3$, and S. Wallon$^2$
%
%
\vspace{.3cm}\\
%
1- CPhT, {\'E}cole Polytechnique, CNRS, 91128 Palaiseau, France \\
%
\vspace{.1cm}\\
2- LPT, Universit{\'e} Paris-Sud, CNRS, 91405 Orsay, France \\
\vspace{.1cm}\\
3- Soltan Institute for Nuclear Studies, Warsaw, Poland\\
}

\maketitle

\begin{abstract}
We discuss the production of two pion pairs in photon collisions at high energies as it can take place in ultraperipheral collisions at hadron colliders such as the LHC. We calculate the according matrix elements in $k_T$-factorization and discuss the possibility to reveal the existence of the perturbative Odderon by charge asymmetries. 
\end{abstract}

\section{Introduction}

In the context of hadronic interactions, events with rapidity gaps are dominated by the exchange of a $C$-even color singlet state, called the Pomeron. Within perturbative QCD the Pomeron is built by the exchange of two gluons at lowest order. While the Pomeron has acquired a good standing, the status of its $C$-odd partner -- the Odderon -- is less safe. Although it is needed {\it e.g.} to describe properly  the different behaviors of $pp$ and $\bar p p$ elastic cross sections \cite{LN}, it still evades confirmation in the perturbative regime, where, again at lowest order, it can be described by the exchange of three gluons in the color singlet state.

The main reason lies in its smaller exchange amplitude in comparison to the Pomeron exchange such that in the cross section, obtained after squaring the sum of both amplitudes, the Pomeron amplitude squared dominates. In this contribution we present results of our study \cite{Pire:2008xe} of a charge asymmetry in the  production of two pion pairs in photon-photon collisions where that Pomeron squared part vanishes. This observable is thus linearly sensitive to the Odderon contribution.

In the present analysis we deal with the hard Pomeron and the hard Odderon 
exchanges, {\it i.e.} both treated within  perturbative QCD. 
This approach can be confronted with a description of the 
Pomeron-Odderon interference based on soft, non-perturbative 
physics and developed in 
Refs.~\cite{Ginzburg:2002zd}. The experimental observation  of the P-O 
interference effects will thus shed a light on the important question 
which of the above mechanisms is more appropriate for the description of data.

\section{Kinematics, amplitudes and GDAs}

\begin{figure}[t]
\centerline{\includegraphics[height=4.9cm]{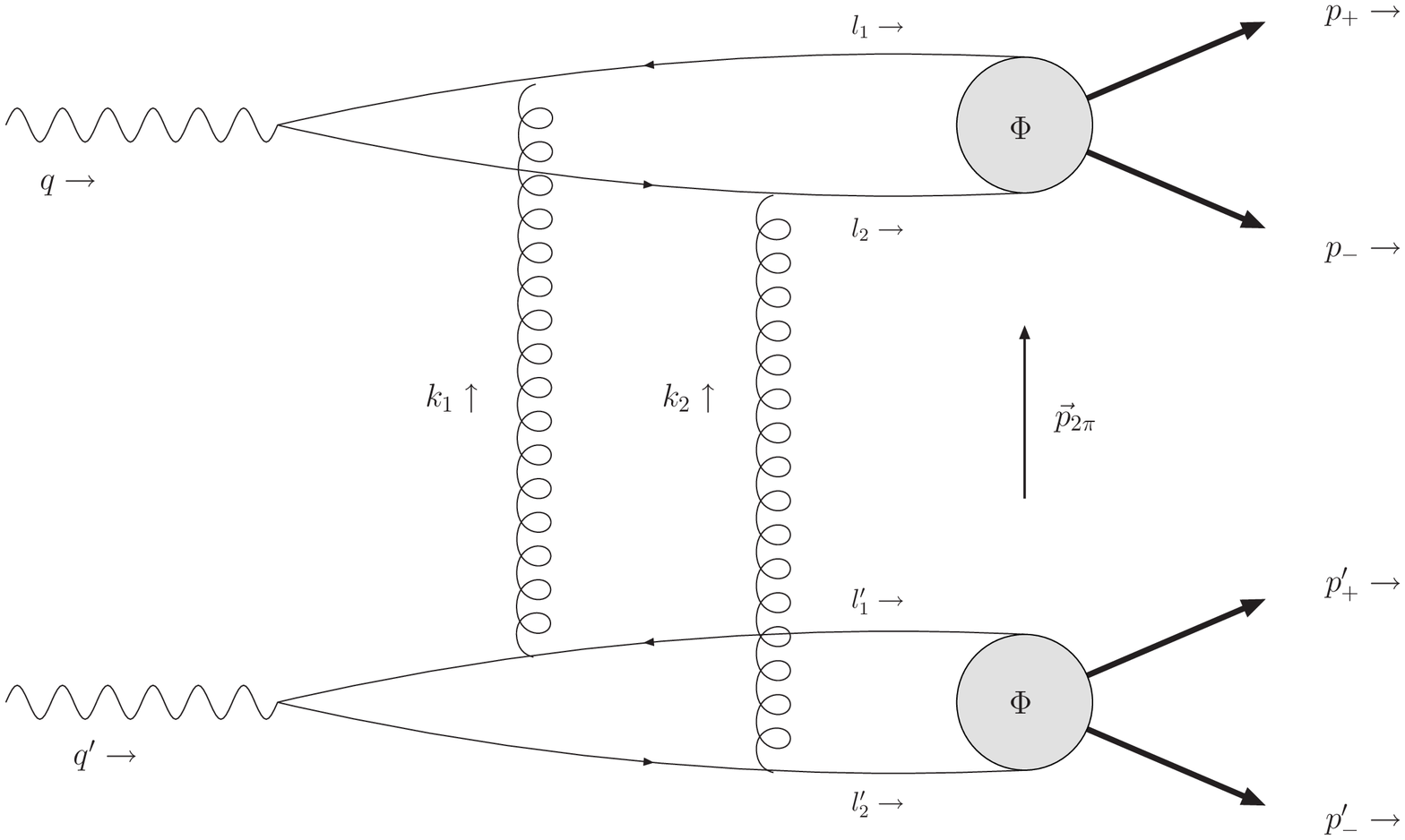}}
\caption{{\protect\small Kinematics of the reaction $\gamma \gamma \to \pi^+ \pi^-\;\; \pi^+ \pi^-$ in a sample Feynman diagram of the two gluon exchange process.}}
\label{fig:1}
\end{figure}

Figure~\ref{fig:1} shows a sample diagram of the process under consideration. We consider large $\gamma\gamma$ energies such that the amplitude can be expressed in terms of two impact factors convoluted over the transverse momenta of the exchanged gluons. The impact factors are universal and consist of a perturbative part -- describing the transition of a photon into a quark-antiquark pair -- and a non-perturbative part, the two pion generalized distribution amplitude (GDA) parametrizing the quark-antiquark hadronization into the pion pair. These 
GDAs~[4-8]
 which are functions of the longitudinal momentum fraction $z$ of the quark,  of
the angle $\theta$ (in the rest frame of the pion pair) and of the invariant mass $m_{2\pi}$ of the pion system are the only but nevertheless essential phenomenological inputs. 
In principle, they have to be extracted from experiments but this is a very challenging task and has not been done so far. However, after an expansion in Gegenbauer polynomials $C_n^m(2z-1)$ and in Legendre polynomials $P_l(\beta\cos\theta)$ (where $\beta=\sqrt{1-4m_\pi^2/m_{2\pi}^2}$) \cite{Polyakov:1998ze}, it is believed that only the first terms give a significant contribution:
\begin{eqnarray*}
  \Phi^{I=1} (z,\theta,m_{2\pi}) &=& 6z\zbar\beta f_1(m_{2\pi}) \cos\theta ,\\
  \Phi^{I=0} (z,\theta,m_{2\pi}) &=& 5z\zbar(z-\zbar)\left[-\frac{3-\beta^2}{2}f_0(m_{2\pi})+\beta^2f_2(m_{2\pi})P_2(\cos\theta)\right],
\end{eqnarray*}
where $f_1(m_{2\pi})$ can be identified with the electromagnetic pion form factor $F_\pi(m_{2\pi})$. 
For the $I=0$ component we use different models. The first model follows Ref.~\cite{Hagler:2002nh} and expresses the functions $f_{0/2}$ in terms of the Breit-Wigner amplitudes of the according resonances.
A second model has been elaborated in Ref.~\cite{Warkentin:2007su} and interprets the functions $f_{0/2}$ as corresponding Omn\`es functions for $S-$ and $D-$waves constructed by dispersion relations from the phase shifts of the elastic pion scattering. 
It has been argued \cite{Warkentin:2007su,Ananthanarayan:2004xy} that the actual phases of the GDA might be closer to the phases  $\delta_{T,l}$ of the corresponding $T$ matrix elements $\frac{\eta_l e^{2i\delta_l}-1}{2i}$, where $\eta_l$ is the inelasticity factor. The third model for the $I=0$ component of the GDA takes this into account by using the technique of model 2 with these phases $\delta_{T,l}$ of the $T$ matrix elements. Indeed, measurements at HERMES \cite{Airapetian:2004sy} do not observe a resonance effect at the $f_0$-mass, but concerning the $f_2$ both phases ($\delta_2$ and $\delta_{T,2}$) are compatible with data \cite{Warkentin:2007su}. Having this in mind, we consider also a fourth model -- a mixed description with the $f_0$ contribution from model 3 and the $f_2$ contribution from model 2. 

\section{Charge asymmetries and rates}

The key to obtain an observable which linearly depends on the Odderon amplitude is the orthogonality of the $C$-even GDA (entering the Odderon process) and the $C$-odd one (entering the Pomeron process) in the space of  Legendre polynomials in $\cos\theta$. Due to an additional multiplication by $\cos\theta$ before the angular integration only the interference term survives. We have to do this for both the pion pairs. Moreover we integrate over the invariant mass of one of the pion pairs to reduce the complexity of our observable. Finally, we define the charge asymmetry in the following way:
\begin{gather*}
 \hat{A}(t,m_{2\pi}^2;m_{\rm min}^2,m_{\rm max}^2) = \frac{\int_{m_{\rm min}^2}^{m_{\rm max}^2} dm_{2\pi}'^2\int\cos\theta\,\cos\theta'\,d\sigma(t,m_{2\pi}^2,m_{2\pi}'^2,\theta,\theta')}{\int_{m_{\rm min}^2}^{m_{\rm max}^2} dm_{2\pi}'^2\int\,d\sigma(t,m_{2\pi}^2,m_{2\pi}'^2,\theta,\theta')}   \\
= \frac{\int_{m_{\rm min}^2}^{m_{\rm max}^2} dm_{2\pi}'^2\int_{-1}^1d\cos\theta\int_{-1}^1d\cos\theta'\;2\cos\theta\,\cos\theta'\,{\rm Re}\left[\mathcal{M}_\pom(\mathcal{M}_\odd+\mathcal{M}_{\gamma})^*\right]}{\int_{m_{\rm min}^2}^{m_{\rm max}^2} dm_{2\pi}'^2\int_{-1}^1d\cos\theta\int_{-1}^1d\cos\theta'\,\left[\left|\mathcal{M}_\pom\right|^2+\left|\mathcal{M}_\odd+\mathcal{M}_{\gamma}\right|^2\right]}
. \label{eq:ahat}
\end{gather*}

\begin{figure}
  \centering
  \includegraphics[width=6cm]{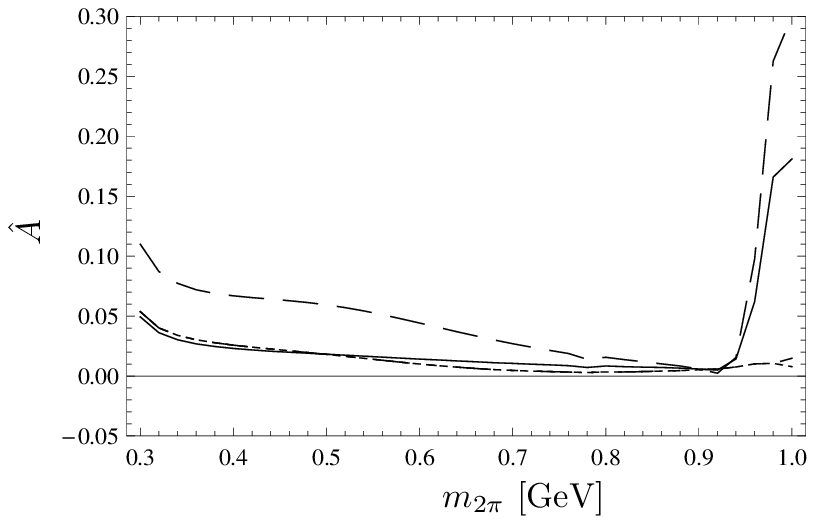}\hspace{10mm}
  \includegraphics[width=6cm]{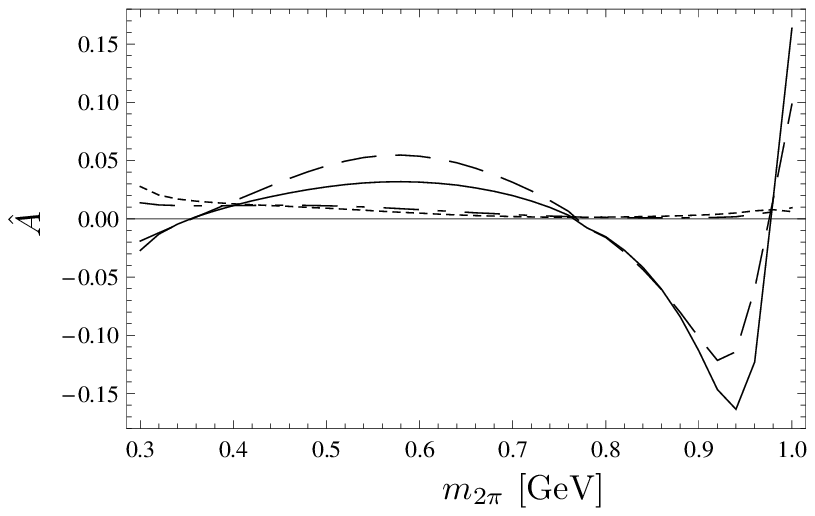}
  \caption{Asymmetry $\hat{A}$ at $t=-1\,\gev^2$ for model 1 (solid), 2 (dashed), 3 (dotted), and 4 (dash-dotted) -- model 3 and 4 are nearly on top of each other. Left column has $m_{\rm min}=.3\,\gev$ and $m_{\rm max}=m_\rho$, while right column has $m_{\rm min}=m_\rho$ and $m_{\rm max}=1\,\gev$. 
}
\label{fig:asymplot1}
\end{figure}

An analytic calculation of the Odderon matrix element would demand the notion of analytic results for two-loop box diagrams, whose off-shellness for all external legs is different. With the techniques currently available such a calculation is not straightforward and we chose to rely on a numerical evaluation by Monte Carlo methods. In particular we make use of a modified  version of {\sc Vegas} as it is provided by the {\sc Cuba} library \cite{Hahn:2004fe}. The result for $\hat{A}$ is shown in Fig.~\ref{fig:asymplot1} where we took two different choices for the integrated region of the invariant mass of the two pions system. Since our framework is only justified for $m_{2\pi} ^2 < -t$, (in fact strictly speaking, one even needs $m_{2\pi}^2 \ll -t$ ), we keep $m_{2\pi}$ below 1\,GeV. 

\begin{figure}
  \begin{center}
\includegraphics[width=12cm]{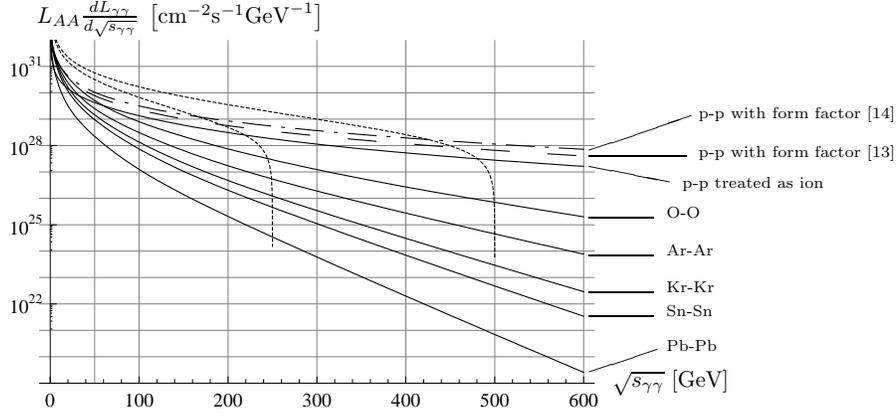} 
  \end{center}
  \caption{Effective $\gamma\gamma$ luminosities for the collision of p-p based on Ref.~\cite{Drees:1988pp} (dash-dotted) and Ref.~\cite{Nystrand:2004vn} (dashed). The results using the parametrization of Ref.~\cite{Cahn:1990jk} for ions are given by solid lines for p-p,
${\rm O}^{8}_{16}$-${\rm O}^{8}_{16}$,
${\rm Ar}^{18}_{40}$-${\rm Ar}^{18}_{40}$,
${\rm Kr}^{36}_{84}$-${\rm Kr}^{36}_{84}$,
${\rm Sn}^{50}_{120}$-${\rm Sn}^{50}_{120}$,
${\rm Pb}^{82}_{208}$-${\rm Pb}^{82}_{208}$ from top to bottom. For ions we used the average luminosities as given in Ref.~\cite{Brandt:2000mu}, for proton we used $L_{pp}=10^{34}\,{\rm cm}^{-2}{\rm s}^{-1}$. For comparison also effective $\gamma\gamma$ luminosities at the ILC are given for $\sqrt{s_{e^+e^-}}=250\,{\rm GeV}$ and $\sqrt{s_{e^+e^-}}=500\,{\rm GeV}$ (both as dotted lines).}
  \label{fig:lumi}
\end{figure}

\begin{figure}
  \centering
  \psfrag{sggingev}{$\begin{matrix}\sqrt{\smin}\\{\rm [GeV]}\end{matrix}$}
  \psfrag{events}{number of events}
  \includegraphics[width=10cm]{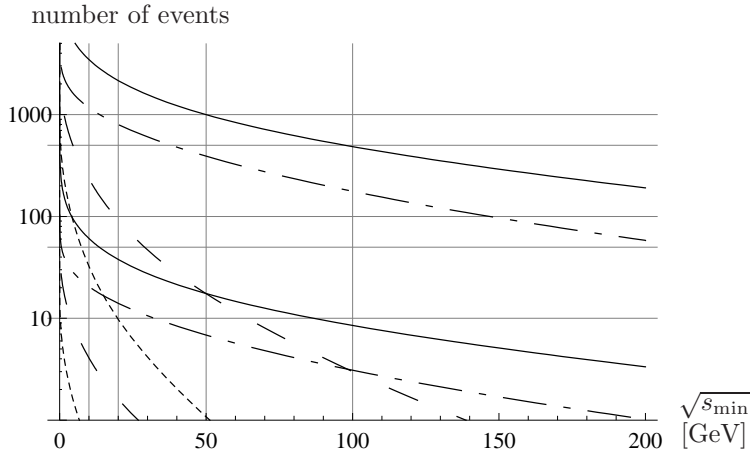}
  \caption{Rate of production of two pion pairs in ultraperipheral collisions for $\tmin=-1\,\gev^2$ in dependence on the lower cut $\smin$ given in `events per month' in case of ions, and `events per six months' in case of protons which in both cases correspond to one year of running of LHC. The solid line displays the result for p-p collision using Ref.~\cite{Drees:1988pp}, the dashed-dotted that for protons treated as heavy ions, the dashed one that for Ar-Ar collisions, and the dotted line that for Pb-Pb collisions. Also the much smaller rates coming from the Odderon exchange are shown (with the same dashing).}
\label{fig:rates}
\end{figure}

To answer the question whether it is possible to measure this asymmetry, we need to discuss the two main issues. First, one has to convolute the $\gamma\gamma$ cross section obtained by our calculations with the effective photon flux at a collider, {\it e.g.} the LHC. As we discuss in Ref.~\cite{Pire:2008xe} in detail, the most recent review on this topic \cite{Baltz:2007kq} presents an overview of photon fluxes for different colliding hadrons which is flawed by an  inconsistency in the underlying hadron-hadron luminosities. Therefore, we show a consistent comparison for the design luminosities of different colliding particles in Fig.~\ref{fig:lumi}. For the proton case we display the different luminosities based on either the proton form factor \cite{Drees:1988pp, Nystrand:2004vn} or on the asymptotic formula for large nuclei -- both leading to sizeable rates at the LHC as can be seen in Fig.~\ref{fig:rates}.

The second important issue is the background. In contrast to many other processes studied in ultraperipheral collisions, pions are produced by pure QCD processes as well. Although heavy ions would offer the possibility to easily trigger on ultraperipheral collisions by detecting neutrons from giant dipole resonances in the Zero Degree Calorimeters, the rates that can be read off from Fig.~\ref{fig:rates} are rather low. 
Regarding the cross section, the p-p mode is highly preferred, but lacks 
the possibility of easy triggering. Still, one could rely 
on the fact that ultraperipheral processes are strongly peaked at low $t$,
contrarily to the flatter depence of the  Pomeron induced ones \cite{Piotrzkowski:2000rx}.


\section{Conclusion}

We have shown that in production of pion pairs in $\gamma\gamma$ collisions the charge asymmetry which is linearly dependent on the Odderon amplitude is sizeable and hence offers the possibility to observe the perturbative Odderon in ultraperipheral collisions at the LHC. The concrete values are GDA-model dependent. HERMES measurements of two pion  electroproduction 
\cite{Airapetian:2004sy} disfavor models with a strong $f_0$ coupling to  
the $\pi^+ \pi^-$ state but to our minds higher statistics data, which  
may come from a JLab experiment at 12\,GeV, are needed  
before  a definite conclusion.

\section*{Acknowledgments}
We acknowledge discussions with Mike Albrow, Gerhard Baur, M{\'a}t{\'e} Csan{\'a}d, David d'En\-terria, Bruno Espagnon, Spencer Klein, Joakim Nystrand, and Rainer Schicker.
This work is supported in part by the Polish Grant N202 249235, the French-Polish scientific agreement Polonium, by the grant ANR-06-JCJC-0084 and by the ECO-NET program, contract 12584QK.



\begin{footnotesize}



\end{footnotesize}


\end{document}